\documentclass[10pt,letterpaper]{article}%
\usepackage{opex3,amsmath,amssymb}
\usepackage{graphicx}
\usepackage{epsfig}
\usepackage{verbatim}
\usepackage{amsmath}
\usepackage{amsfonts}
\usepackage{amssymb}%
\providecommand{\U}[1]{\protect\rule{.1in}{.1in}}
\begin{document}

\title{Photon pair-state preparation with tailored spectral properties by spontaneous
four-wave mixing in photonic-crystal fiber}
\author{K. Garay-Palmett,$^{1 *}$ H. J. McGuinness,$^{2}$ Offir Cohen,$^{3}$ J. S.
Lundeen,$^{3}$\\R. Rangel-Rojo,$^{1}$ A. B. U'Ren,$^{1}$ M. G.
Raymer,$^{2}$ C. J. McKinstrie,$^{4}$\\ S. Radic,$^{5}$  and I. A.
Walmsley$^{3}$}
\address{ $^1$ Departamento de Optica, Centro de Investigaci\'on
Cient\'{\i}fica y de Educaci\'on Superior de
Ensenada, Apartado Postal 2732, Ensenada, BC 22860, Mexico.\\
$^2$ Department of Physics, University of Oregon, Eugene, OR 97403, United States. \\
$^3$ Clarendon Laboratory, Oxford University, Parks Road, Oxford,
OX1.
3PU, United Kingdom \\
$^4$ Bell Laboratories, Alcatel--Lucent, Holmdel, NJ 07733, United States. \\
$^5$ Department of Electrical and Computer Engineering, University
of California at San Diego, La Jolla, CA 92093, United States.}

\email{kgaray@cicese.mx}

\begin{abstract}
We study theoretically the generation of photon pairs by
spontaneous four-wave mixing (SFWM) in photonic crystal optical
fiber. We show that it is possible to engineer two-photon
states with specific spectral correlation (``entanglement'')
properties suitable for quantum information processing
applications. We focus on the case exhibiting no spectral
correlations in the two-photon component of the state, which we
call factorability, and which allows heralding of single-photon
pure-state wave packets without the need for spectral post
filtering. We show that spontaneous four wave mixing exhibits a
remarkable flexibility, permitting a wider class of two-photon
states, including ultra-broadband, highly-anticorrelated
states.
\end{abstract}






\ocis{(190.4380)Nonlinear optics, four-wave mixing; (270.0270)
Quantum optics.}



\section{Introduction}

Quantum optical technologies require photon states with specific
spectral properties. For example, quantum information processing
using linear optics is based upon the availability of pure-state
single-photon wavepackets \cite{kok07}. Single-photon wavepackets
can be prepared using pair generation by means of spontaneous
parametric downconversion (PDC), or by spontaneous four wave mixing
(SFWM) \cite{harris67}. In both cases individual photons are
heralded by the detection of their siblings. The prepared photons
will not normally be in pure states unless special care is taken to
remove all correlations in every degree of freedom of the photon
pairs, i.e. to make the two-photon state factorable\cite{uren05}. If
this is not done, the heralded photons will be in mixed states, and
therefore unsuitable for use in quantum logic gates, which rely on
Hong-Ou-Mandel (HOM) interference between independent photons.
Typically, photon pairs generated in spontaneous processes exhibit
significant spectral and spatial correlations due to the energy and
momentum conservation constraints that are typical in parametric
nonlinear optics. Spatial correlations may be minimized by use of
guided-wave configurations, such as those exploiting nonlinear
waveguides and optical fibers\cite{raymer05,
uren04,banaszek99,fan06,rarity05,fan2005,li04}.

Spectral correlations are, however, more difficult to eliminate. It
is possible to eliminate all correlations in this degree of freedom
for photon pairs generated by means of PDC, as first shown in Ref.
\cite{grice01}, and later extended in \cite{uren05}. This is
achieved by the method of group-velocity matching using a broadband
pump pulse. In this paper we generalize this method to the case of
SFWM, by engineering the group velocities of the sideband photons
using photonic crystal fibers (PCFs). PCFs are comprised of a solid
silica core surrounded by a silica cladding containing a regular
array of air holes. This leads to an exceptionally high
core-cladding index contrast, creating strong waveguide dispersion
that can be tailored for a wide-range of applications
\cite{russell03}.

Spontaneous four wave mixing occurs in single-mode fibers with a third-order
optical nonlinearity. In this process, two pump photons are scattered from one
or two distinct pump fields into a pair of fields, labeled signal and idler,
which are spectrally and/or polarization distinct. The single-mode waveguide
geometry leads to the suppression of correlations in the transverse momentum
degree of freedom, as well as suppression of mixed spectral-transverse
momentum correlations. Thus, by imposing appropriate group velocity matching
constraints on the SFWM process in a single-mode fiber, it becomes possible to
eliminate correlations in \textit{all} degrees of freedom, resulting in
factorable two-photon states.

Previous experimental work has explored this type of source in a number of
geometries including standard single-mode fiber and PCF. Photon number
correlations between signal and idler fields were first observed by pumping in
the anomalous dispersion region \cite{fiorentino02}, although there was a
significant background of noise photons generated by Raman scattering of the
pump light \cite{stolen2003}. Spontaneous Raman scattering generates photons
shifted to the red of the pump wavelength by up to 50 THz, corresponding to
the frequency of an optical phonon in glass. Thermal population of the phonons
gives rise to scattering over a broad frequency range, and the process does
not require phase matching, so it occurs at all pump wavelengths. Photonic
crystal fibers allow the signal and idler fields to be widely separated in
frequency from the pump, so that Raman noise at the lower-frequency photon
wavelength is greatly reduced \cite{fan06,rarity05}. Phase-matching of SFWM
with widely separated sidebands is enabled by pumping in the normal dispersion
region, which requires the favorable dispersion properties of strongly guided
waves \cite{jiang06,harvey03, yu95}.

These experiments did not, however, address the requirement of
purity for generating heralded single-photon states. In particular,
the photons generated in those experiments would require tight
spectral filtering to eliminate correlations between signal an idler
frequencies prior to heralding. Using such tight filtering ensures
high-visibility HOM interference between photons from separate
sources, but greatly reduces the count rates. To avoid the need for
filtering, the SFWM process must produce a state whose two-photon
component can be written as the product of the signal state and the
idler state, i.e.,$|\psi\rangle=|0\rangle_{s}|0\rangle_{i}
+\kappa|1\rangle _{s}|1\rangle_{i}$, in the photon number basis with
$s,i$ indicating signal and idler modes respectively. We call this
property ``factorability,'' which corresponds to the absence of
correlations between the frequencies (and momenta) of the idler and
signal photons. In physical terms, factorability implies that no
information about, say, the idler photon (apart from its existence)
can be extracted from the detection of the signal photon, or
vice-versa. To date, such factorable states have been produced only
in PDC at a single wavelength using a particular crystal having
special dispersion properties \cite{mosley07}, and other techniques
could in principle be used to extend possible operation wavelengths
\cite{raymer05,walton04,torres05,kuzucu05,uren03,uren06}. We show
here that factorability can be achieved at a much wider range of
wavelengths by using PCF.

In addition to factorability, PCF enables the production of a wide range of
spectral correlations in the two-photon component of the state. Two extremes
are possible, a spectrally correlated state and a spectrally anti-correlated
state. The former is a resource for quantum-enhanced quantum positioning
\cite{giovanetti01}, whereas the latter is of importance for applications
relying on time-of-arrival differences between two optical modes, such as
optical coherence tomography \cite{nasr03}. We will show that the
anti-correlated case can be made ultra-broadband by tailoring the higher-order
dispersion. Thus we see that the ability to engineer the dispersion in PCF
leads to a flexible system that can generate states in two widely disparate
regimes (factorable and highly correlated), and indeed in a very general class
of intermediate regimes.

In all these cases, the light is produced in well-defined fiber modes, which
is convenient for integration with waveguide devices. This holds the promise
of combining many quantum logic gates using integrated optics, leading to the
possibility of scalable quantum optical devices.

\section{Spontaneous four wave mixing theory}

In this paper we study SFWM in a single-mode optical fiber with a third-order
nonlinear susceptibility $\chi^{(3)}$. In this process, two-photons from pump
fields $E_{1}$ and $E_{2}$ are jointly annihilated to create a photon pair
comprised of one photon in the signal mode, $\hat{E}_{s} $, and one photon in
the idler mode $\hat{E}_{i}$. We assume that all fields propagate in the
fundamental spatial mode of the fiber. This assumption is justified if the
fiber core radius is small enough that it only supports the fundamental mode,
or alternatively if only this fundamental mode is excited. Following a
standard perturbative approach \cite{mandel}, the two-photon state produced by
spontaneous four-wave mixing in an optical fiber of length $L$ can be shown to
be given by \cite{chen05,chen07arxiv}%

\begin{align}
\label{eq: state} &  |\Psi\rangle= |0\rangle_{s}|0\rangle_{i} +\kappa\int\int
d\omega_{s} d\omega_{i}F\left(  \omega_{s},\omega_{i}\right)  \left|
\omega_{s}\right\rangle _{s} \left|  \omega_{i}\right\rangle _{i}.
\end{align}


Here, $\kappa$ is a constant which represents the generation efficiency
(linearly proportional to the fiber length, electric field amplitude for each
of the pump fields and dependent on the relative polarizations of the pump and
created pair fields) and $F\left(  \omega_{s},\omega_{i}\right)  $ is the
joint spectral amplitude function (JSA), which describes the spectral
entanglement properties of the generated photon pair%

\begin{align}
\label{eq: JSA}F\left(  \omega_{s},\omega_{i}\right)   &  = \int
d\omega^{\prime}\alpha_{1}\left(  \omega^{\prime}\right)  \alpha_{2} \left(
\omega_{s}+\omega_{i}-\omega^{\prime}\right) \nonumber\\
&  \times\mbox{sinc}\left[  \frac{L}{2} \Delta k\left(  \omega^{\prime}%
,\omega_{s},\omega_{i}\right)  \right]  \mbox{exp} \left[  i\frac{L}{2}\Delta
k\left(  \omega^{\prime},\omega_{s},\omega_{i}\right)  \right]  ,
\end{align}

\noindent which is given in terms of the pump spectral amplitudes
$\alpha_{1,2} (\omega)$, and the phase mismatch function $\Delta k\left(
\omega_{1},\omega_{s},\omega_{i}\right)  $, that in the case where the two
pumps, signal and idler are co-polarized, is given by

%

\begin{equation}
\Delta k\left(  \omega_{1},\omega_{s},\omega_{i}\right)  =k\left(  \omega
_{1}\right)  +k\left(  \omega_{s}+\omega_{i}-\omega_{1}\right)  -k\left(
\omega_{s}\right)  -k\left(  \omega_{i}\right)  -\left(  \gamma_{1}%
P_{1}+\gamma_{2}P_{2}\right)  , \label{eq: delk}%
\end{equation}
which includes a self/cross-phase modulation contribution for the two pumps
with peak powers $P_{1}$ and $P_{2}$, characterized by the nonlinear
parameters $\gamma_{1}$ and $\gamma_{2}$, which depend on specific fiber used
and pump wavelength\cite{agrawal2007,mckinstrie06} . The energy conservation
constraint is apparent in the argument of the second term of the phase
mismatch (see Eq.(\ref{eq: delk})). A ``factorable'' state is defined to be a
state for which $F\left(  \omega_{s},\omega_{i}\right)  $ is equal to a
product of two functions, $F\left(  \omega_{s},\omega_{i}\right)
=S(\omega_{s})I(\omega_{i})$, where the functions $S(\omega)$ and $I(\omega)$
depend only on the signal and idler frequencies respectively.

By making use of a linear approximation for the phase mismatch, in addition to
modeling $\alpha_{1,2}(\omega)$ as Gaussian functions with bandwidth
$\sigma_{1,2}$ respectively, it is possible to obtain an expression for the
joint spectral amplitude in closed analytical form. Expanding $k\left(
\omega_{\mu}\right)  $ in a first-order Taylor series about frequencies
$\omega_{\mu}^{0}$ for which perfect phase-matching is attained (where
$\mu=1,2,s,i$), and defining the detunings
$\nu_{s}=\omega_{s}-\omega_{s}^{0}$ and $\nu_{i}=\omega_{i}-\omega_{i}^{0}$,
the approximate phase mismatch $\Delta k_{lin}$ is defined by%

\begin{equation}
\label{eq: ldelk}L \Delta k_{lin} = L\Delta k^{\left(  0\right)  }+T_{s}
\nu_{s} +T_{i} \nu_{i},
\end{equation}

\noindent where $\Delta k^{(0)}$, given by Eq.(\ref{eq: delk}) evaluated at
the frequencies $\omega_{\mu}^{0}$, must vanish to guarantee phase-matching at
these center frequencies. The coefficients $T_{\mu}$ are given by $T_{\mu
}=\tau_{\mu}+\tau_{p}\sigma_{1}^{2}/(\sigma_{1}^{2}+\sigma_{2}^{2})$, where
$\tau_{\mu}$ represent group-velocity mismatch terms between the pump centered
at frequency $\omega_{2}^{0}$ and the generated photon centered at the
frequency $\omega_{\mu}^{0}$, and $\tau_{p}$ is the group velocity mismatch
between the two pumps%

\begin{align}
\label{eq: taus}\tau_{\mu}= L\left[  k_{2}^{(1)}\left(  \omega_{2}^{0}\right)
- k_{\mu}^{(1)}\left(  \omega_{\mu}^{0}\right)  \right]  ,\nonumber\\
\tau_{p} = L\left[  k_{1}^{(1)}\left(  \omega_{1}^{0}\right)  - k_{2}%
^{(1)}\left(  \omega_{2}^{0}\right)  \right]  ,
\end{align}

\noindent written in terms of $k_{\mu}^{(n)}\left(  \omega\right)
=d^{n}k_{\mu}/d\omega^{n} |_{\omega=\omega_{\mu}^{0}}$ . Note that this
approach requires \textit{a priori} knowledge, for given pump fields, of the
signal and idler frequencies ($\omega_{s}^{0}$ and $\omega_{i}^{0}$) at which
perfect phase-matching is achieved. These frequencies can be determined by
solving (for example numerically) the condition $\Delta k^{(0)}=0$ (see
Eq.(\ref{eq: delk})). It can be shown that within the linear approximation,
the integral in Eq. (\ref{eq: JSA}) can be carried out analytically, yielding%

\begin{equation}
\label{eq: JSANUS}F_{lin}\left(  \nu_{s},\nu_{i}\right)  =\alpha\left(
\nu_{s},\nu_{i}\right)  \phi\left(  \nu_{s},\nu_{i}\right)  ,
\end{equation}

\noindent where the pump envelope function $\alpha\left(  \nu_{s},\nu
_{i}\right)  $ is derived from the pump spectral amplitudes for the two
individual pump fields through the integral in Eq. (\ref{eq: JSA}), and is
given by

%

\begin{equation}
\label{eq:PE}\alpha\left(  \nu_{s},\nu_{i}\right)  =\mbox{exp}\left[
-\frac{\left(  \nu_{s}+\nu_{i} \right)  ^{2}}{\sigma_{1}^{2}+\sigma_{2}^{2}%
}\right]  ,
\end{equation}

\noindent and where $\phi\left(  \nu_{s},\nu_{i}\right)  $ describes the
phase-matching properties in the fiber. For degenerate pumps (where
$\alpha_{1}(\omega)=\alpha_{2}(\omega)$), it may be shown that the
phase-matching function is given by%

\begin{equation}
\label{eq: PhiNsNiI}\phi\left(  \nu_{s},\nu_{i}\right)  = \mbox{sinc}\left[
\frac{L\Delta k_{lin}}{2} \right]  \exp\left[  i\frac{L\Delta k_{lin}}{2}
\right]  ,
\end{equation}

\noindent where $L\Delta k_{lin}$ is given in Eq. (\ref{eq: ldelk}) with
$\tau_{p}=0$. For non-degenerate pumps, $\tau_{p}\neq0$ and $\phi\left(
\nu_{s},\nu_{i}\right)  =\Phi(B;L\Delta k_{lin})$, with%

\begin{align}
\label{PMF1}\Phi(B;x) = M\sqrt{\pi} B \exp(-B^{2} x^{2}) \left[
\mbox{erf}\left(  \frac{1}{ 2B}-i B x\right)  +\mbox{erf}(i B x)\right]  ,
\end{align}

\noindent where erf($z$) is the error function, the parameter $B$ is defined
as $B=\left(  \sigma_{1}^{2}+\sigma_{2}^{2}\right)  ^{1/2}/\left(  \sigma_{1}
\sigma_{2} \tau_{p} \right)  $, and $M$ is a normalization coefficient. In the
present paper, we concentrate on the important class of factorable states for
which $F\left(  \omega_{s},\omega_{i}\right)  =S\left(  \omega_{s}\right)
I\left(  \omega_{i}\right)  $.

\begin{figure}[t]
\begin{center}
\centering\includegraphics[width=13 cm]{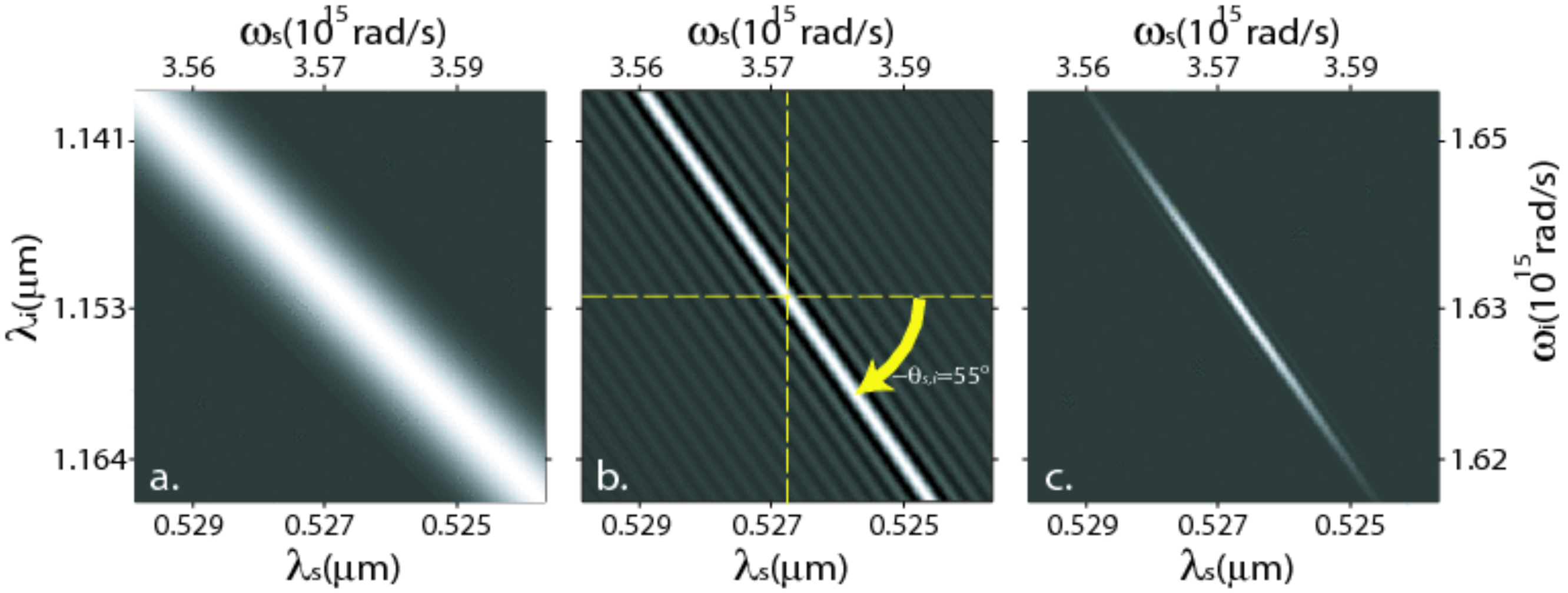}
\end{center}
\caption{(a) Pump envelope function $\alpha(\omega_{s},\omega_{i})$ for a
fiber characterized by $r=0.67\mu$m, $f=0.52$ and $L=30$cm. (b) Phasematching
function $\phi\left(  \nu_{s},\nu_{i}\right)  $; for a relatively small region
of \{$\omega_{s},\omega_{i}$\} space the phase-matching function contours are
essentially straight lines, with slope $\theta_{si}=-\arctan(T_{s}/T_{i})$
(with $T_{\mu}$ given by Eq. (\ref{eq: taus})). (c) Resulting joint spectral
intensity. }%
\label{fig:PMillus}%
\end{figure}

\section{Phase and group-velocity matching properties of photonic crystal
fibers}

\label{sec: disp} Our analysis focuses on PCFs consisting of a fused silica
core surrounded by silica cladding with a pattern of air holes which remains
constant along the fiber length. This mixture of air and and glass in the
cladding results in an average refractive index that is considerably lower
than that of the core, providing a high dielectric contrast, resulting in
strong optical confinement. This leads to high peak irradiances even for
modest input powers, which enhances nonlinear optical effects such as SFWM. In
addition, the dispersion characteristics of the PCFs can be engineered by
variations of the distribution, size and shape of the air holes surrounding
the core. In particular, it becomes possible to choose the zero dispersion
wavelength(s) (ZDW), to tailor the FWM phase-matching properties
\cite{hansen03} and to design fibers approaching endlessly single-mode
behavior \cite{birks97}.


\begin{figure}[t]
\begin{center}
\centering\includegraphics[width=12cm]{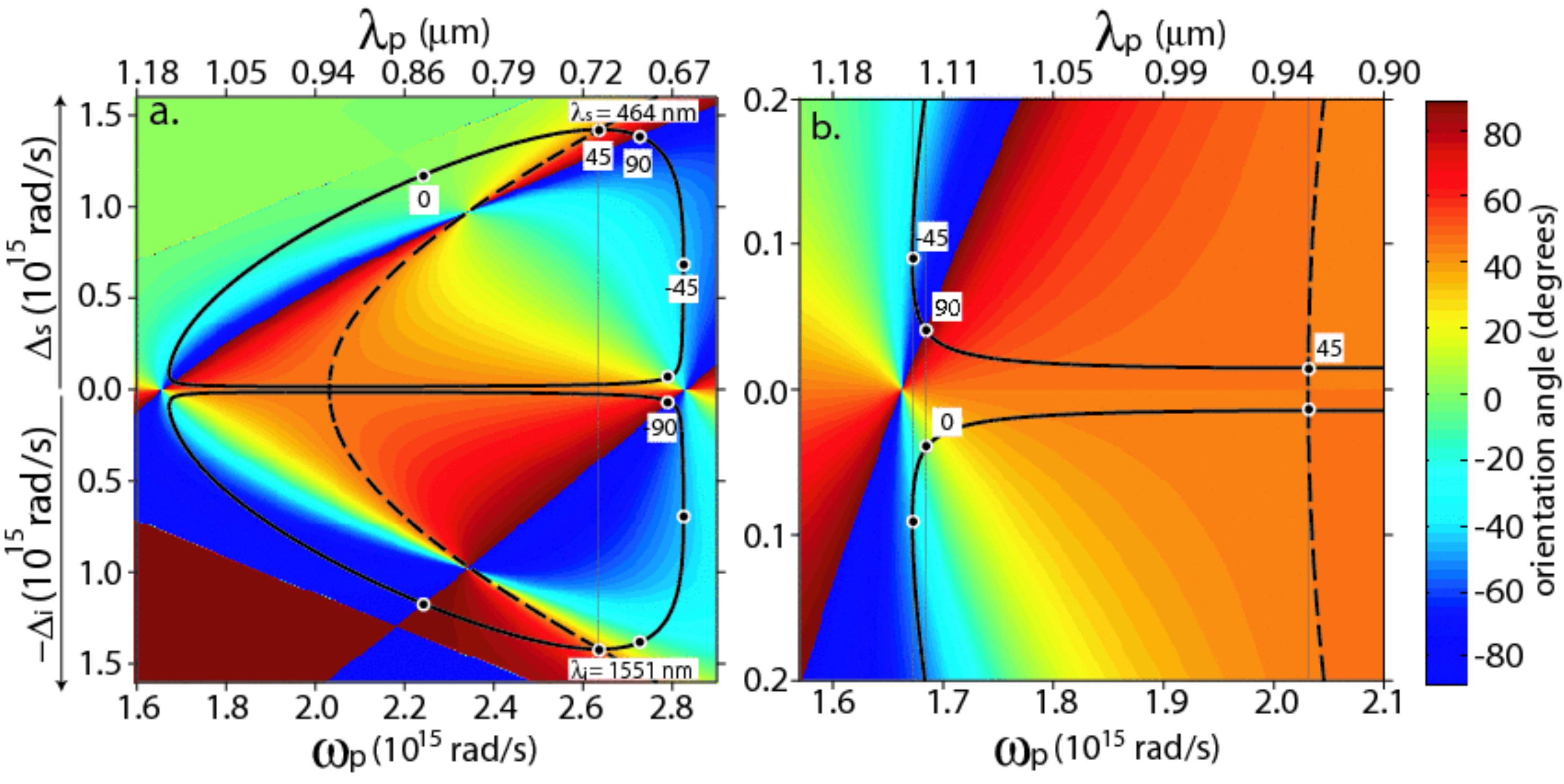}
\end{center}
\par
\caption{(a) Black, solid curve: phase-matching ($\Delta k=0$) contour for
SFWM in the degenerate pump case. Colored background: phase-matching
orientation angle. Black, dashed line: symmetric group velocity matching (GVM)
contour. Along the phase-matching contour we have indicated particular
orientation angles of interest. (b) Close up, near the lower zero group
velocity dispersion frequency.}%
\label{fig: contordk}%
\end{figure}


The FWM phase-matching properties are determined by the fundamental mode
propagation constant, given in terms of the effective refractive index
$n_{eff}$ by $k(\omega)=n_{eff}(\omega)\omega/c$. We adopt a step-index model,
where the core has radius $r$, its index is that of fused silica $n_{s}%
(\omega)$, and the cladding index is calculated as $n_{clad}(\omega
)=f+(1-f)n_{s}(\omega)$, where $f$ is the air-filling fraction. This fiber
dispersion model has been shown to be accurate for $f$ from $0.1$ to $0.9$
according to Ref.~\cite{Wong2005}. In the context of our work, this model
permits a straightforward exploration of the spectral entanglement properties
in $\{r,f\}$ parameter space.

The structure of the SFWM two-photon state, in the degenerate-pump,
co-polarized case (for which $\gamma_{1}=\gamma_{2}=\gamma$) is illustrated in
Fig.~\ref{fig:PMillus}, for a specific fiber with $r=0.67\mu$m, $f=0.52$ and
length $L=30$cm. Fig.~\ref{fig:PMillus}(a) shows the pump envelope function
plotted as a function of $\omega_{s}$ and $\omega_{i}$ (see Eq.(\ref{eq:PE}))
where we have assumed that the pump is centered at $723$nm, has a bandwidth of
$1$nm, the incident power is $5$W and $\gamma=70 \text{km}^{-1} \text{W}^{-1}%
$. Fig.~\ref{fig:PMillus}(b) shows the phasematching function (see
Eq.(\ref{eq: PhiNsNiI})) for this choice of parameters. Note that the
phasematched region forms a strip on $\{\omega_{s},\omega_{i}\}$ space,
oriented at an angle $\theta_{si}=-\arctan(\tau_{s}/\tau_{i})=-55^{\circ}$
with respect to the $\omega_{s}$ axis. Fig.~\ref{fig:PMillus}(c) shows the
resulting joint spectral intensity $|F_{lin}(\omega_{s},\omega_{i})|^{2}$ (see
Eq.(\ref{eq: JSANUS})). It is apparent from Fig.~\ref{fig:PMillus} that the
properties of the two photon state are determined by the i) relative
orientations and ii) widths of the strips representing the phasematching and
pump envelope functions. In this paper we explore how the interplay of the
various design parameters can lead to two-photon states with engineered
spectral entanglement properties.

A plot of the perfect phase-matching contour ($\Delta k(\omega_{p},\omega
_{s},2\omega_{p}-\omega_{s})=0$) for degenerate pumps versus $\omega_{s}$ and
$\omega_{p}$ gives for each pump frequency the expected signal and idler
central frequencies $\omega_{s}^{0}$ and $\omega_{i}^{0}$. Such a
phase-matching contour is illustrated in Fig.~\ref{fig: contordk} for
$r=0.616\mu$m and $f=0.6$, with pump power $30$W and $\gamma=70\text{km}%
^{-1}\text{W}^{-1}$, where the generated frequencies are expressed as
detunings from the pump frequency $\Delta_{s,i}=\omega_{s,i}-\omega_{p}$. Note
that energy conservation implies that $\Delta_{s}=-\Delta_{i}$. PCFs often
have two ZDWs, where one can be as low as 500 nm (in comparison with 1270 nm
for bulk silica). The phase-matching contours take the form of closed loops,
with inner branches near the pump frequency and outer branches that can be
hundreds of nanometers from the pump wavelength. Note that four wave mixing
relying on outer-branch phasematching has been observed in previous work, in
the context of classical non-linear optics~\cite{jiang06}.



For this specific fiber geometry, a continuous pump wavelength range of
approximately $453$ nm exists, in which parametric generation can be observed,
almost completely contained between the two ZDWs ($\lambda_{zd1}=0.668\mu$m
and $\lambda_{zd2}=1.132\mu$m). The power-induced phase modulation terms in
$\Delta k$ split the trivial $\Delta_{s}=\Delta_{i}=0$ branch, leading to
frequency-distinct inner-branch signal/idler frequencies close to the pump
(used for SFWM in \cite{fiorentino02}), whereas the outer branch exhibits
comparatively little dependence on pump power \cite{jiang06,harvey03,yu95}.

In what follows we will concentrate on phase-matching configurations
where the signal and idler frequencies are sufficiently removed from
the pump frequency, or where the pump is orthogonally polarized to
the signal and idler photons, so as to avoid the generation of
background photons by spontaneous Raman scattering at the
lower-frequency photon of the pair \cite{stolen2003}.

From Eq. (\ref{eq:PE}) it is clear that the pump envelope function in
$\{\omega_{s},\omega_{i}\}$ space has contours of equal amplitude which have
negative unit slope in all of $\{\omega_{s},\omega_{i}\}$ space. In contrast,
the contours of the phase-matching function $\phi\left(  \omega_{s},\omega
_{i}\right)  $ are characterized by a slope in $\{\omega_{s},\omega_{i}\}$
space given by $\theta_{si}=-\arctan(T_{s}/T_{i})$ (with $T_{\mu}$ given
according to Eq. (\ref{eq: taus})). The relationship between the slope
$\theta_{\Delta p}$ of the curve at a point in \{$\Delta_{s,i},\omega_{p}$\}
space to the resulting phasematching function in \{$\omega_{s},\omega_{i}$\}
space is $\theta_{\Delta p}=45^{\circ}-\theta_{si}$. For example, $\theta
_{si}=45^{\circ}$ corresponds to zero slope on the contour in Fig.
\ref{fig: contordk}(a) or $\theta_{\Delta p}=0$. The colored background in
Fig.~\ref{fig: contordk} indicates the slope of the phase-matching contour, or
orientation angle, ranging from $\theta_{si}=-90^{\circ}$ to $\theta
_{si}=+90^{\circ}$, indicated in blue and red respectively.

The type of spectral correlations observed in a SFWM two-photon state is
determined in part by the slope $\theta_{si}$ of the phase-matching contour.
If the phase-matching contour is given by a closed loop (which is true for
most fibers, i.e. range of values of $\{r,f\}$, of interest), all
phase-matching orientation angles $\theta_{si}$ are possible, controlled by
the pump frequency. Thus, for certain relative orientations and widths of
these two functions, it becomes possible to generate factorable two-photon
states. It can be shown that a factorable state is possible if%

\begin{equation}
\label{eq: CoDP_i}T _{s}T _{i} \leq0.
\end{equation}

Among those states which fulfil Eq.(\ref{eq: CoDP_i}) those exhibiting a
phasematching angle of $\theta_{si}=45^{\circ}$, or $T_{s}=-T_{i}$, are of
particular interest. For these states, in the degenerate pumps case, a
factorable, symmetric state is guaranteed if%

\begin{equation}
\label{eq: CoDP_ii}2\Gamma\sigma^{2}|T _{s}T _{i}|=1,
\end{equation}

\noindent where $\Gamma\approx0.193$. The condition in Eq. (\ref{eq: CoDP_i})
constrains the group velocities: either $k^{(1)}(\omega_{s})<k^{(1) }%
(\omega_{p})<k^{(1) }(\omega_{i})$, or $k^{(1) }(\omega_{i})<k^{(1) }%
(\omega_{p})<k^{(1) }(\omega_{s})$ must be satisfied. The condition in Eq.
(\ref{eq: CoDP_ii}) constrains the bandwidths. Thus, the region in
$\{\omega_{s},\omega_{i}\}$ space in which factorability is possible is
bounded by the conditions $T_{s}=0$ and $T _{i}=0$.

Note that in the case of PDC (in second-order nonlinear crystals), it is in
general terms challenging to obey Eq. (\ref{eq: CoDP_i}), for it implies that
the pump, say with frequency $\omega_{p}$, must propagate at a higher group
velocity than one of the generated photons at $\omega_{p}/2$, which can be
interpreted as anomalous group-velocity dispersion. In practice, this can be
achieved for certain materials, within a restricted spectral region (usually
with PDC in the infrared) in a type-II process, where the polarization of the
pump is orthogonal to that of one of the generated photons~\cite{mosley07}. In
the case of SFWM, because the frequencies (in the degenerate pump regime) will
in general obey $\omega_{s}<\omega_{p}<\omega_{i}$ or $\omega_{i}<\omega
_{p}<\omega_{s} $, no such anomalous group-velocity dispersion is needed,
making it much more straightforward to fulfill the group-velocity matching
conditions required for factorable photon pair generation.

Once the dispersion relation for a given fiber geometry has been
determined, the next step in designing a factorable photon pair
source consists of identifying the pump wavelengths that satisfy the
required group-velocity matching, amongst those which also satisfy
phase-matching. For this purpose, we refer to the colored background
in Fig. \ref{fig: contordk}. Each color corresponds to a different
phase-matching orientation angle $\theta_{si}$. In particular,
symmetric factorable two-photon states, for which the signal and
idler photons have identical spectral widths, are possible if
$\theta _{si}=45^{\circ}$. In this case symmetric group-velocity
matching $2k_{p}^{(1) }=k_{s}^{(1) }+k_{i}^{(1) }$ (or equivalently
$T_{s}= -T_{i}$) is attained, and the phase-matching contours are
oriented so that its contours have unit slope. The frequency values
that fulfill this condition are represented by the dashed line. The
pump frequency that permits symmetric factorable states can be
determined from the intersection of the phase-matching contour with
the group-velocity matching contour.

Similarly, asymmetric factorable two-photon states, for which the signal and
idler photons have greatly different spectral widths, are possible if
$\theta_{si}=0^{\circ}$ or $\theta_{si}=90^{\circ}$ . This is the case of
asymmetric group-velocity matching, for which $k_{p}^{(1)}=k_{s}^{(1) }$ or
$k_{p}^{(1) }=k_{i}^{(1) }$ (or equivalently $T_{s}=0$ or $T_{i}=0$) is
required. In this case the phase-matching contours are oriented parallel to
the $\omega_{s}$ or $\omega_{i}$ axes. In addition, Eq. (\ref{eq: CoDP_ii})
leads to the condition that $T_{i}\gg1/\sigma$ (for $T_{s}=0$) or $T_{s}%
\gg1/\sigma$ (for $T_{i}=0$).

Both symmetric and asymmetric GVM are possible using a number of different
configurations. We provide examples of each, exploiting co-polarized as well
as cross-polarized SFWM, in the following sections. Note however, that the
examples to be shown do not include all possible source designs.

\section{Co-polarized fields and degenerate pumps: symmetric factorable
states}

\begin{figure}[t]
\begin{center}
\centering\includegraphics[width=11.5cm]{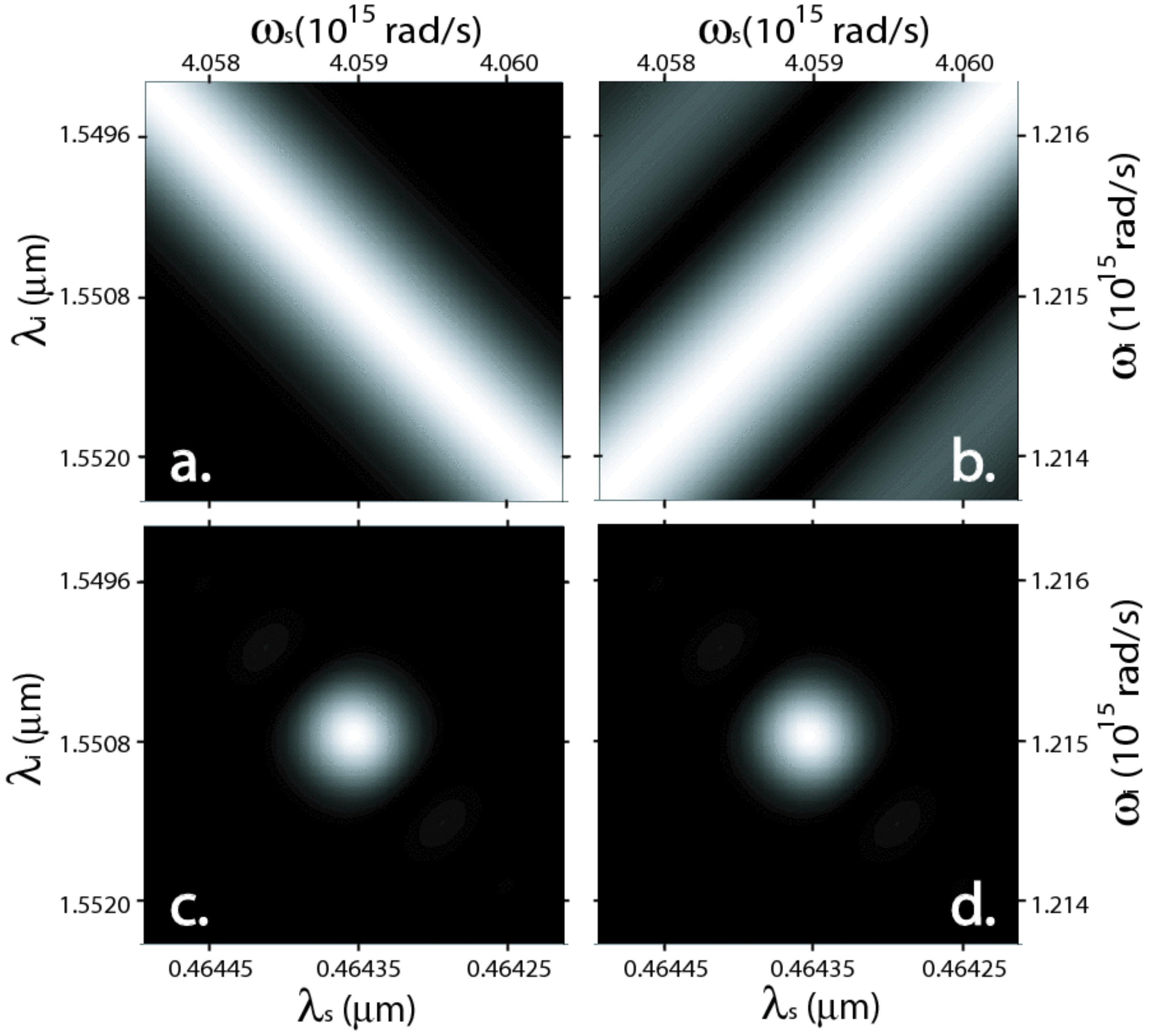}
\end{center}
\caption{Joint spectral intensity (JSI) obtained for the fiber geometry
assumed for Fig. \ref{fig: contordk} ($r=0.616 \mu$m and $f=0.6$) where the
pump central wavelength ($\lambda_{p0}=0.7147\mu$m) is obtained by imposing
simultaneous phase-matching and group-velocity matching. We consider a pump
bandwidth of $0.1$nm and a fiber length of $0.25$m. The function values are
normalized such that white $=1$ and black $=0$. (a) Pump envelope function
$\alpha(\omega_{s},\omega_{i})$. (b) Phase-matching function $\phi\left(
\nu_{s},\nu_{i}\right)  $. (c) Analytic JSI, obtained with approximation from
Eq. (\ref{eq: ldelk}). (d) JSI obtained by numerical integration of Eq.
(\ref{eq: JSA}) .}%
\label{fig:estsime}%
\end{figure}

The synthesis of a symmetric, factorable state is illustrated in Fig.
\ref{fig:estsime} for the same fiber geometry as described above ($r=0.616\mu
$m and $f=0.6$). Figure \ref{fig:estsime}(a) shows the pump envelope function,
for a
single pump centered at $715$ nm with a relatively narrow bandwidth of
$0.1$nm. Figure \ref{fig:estsime}(b) shows the phase-matching function
assuming a fiber length of $25$cm, while Fig. \ref{fig:estsime}(c) shows the
joint spectral intensity $|F_{lin}(\omega_{s},\omega_{i})|^{2}$, exhibiting an
essentially factorable character. For comparison, Fig. \ref{fig:estsime}(d)
shows the joint spectral intensity obtained by numerical integration of Eq.
(\ref{eq: JSA}) using the full dispersion (rather than relying on a linear
approximation), revealing that this case the linear approximation of the
phase-matching condition is in fact an excellent approximation. This source
leads to a numerically-obtained state purity defined as Tr[$\hat{\rho}_{s}%
^{2}$], where $\hat{\rho}_{s}$ is the reduced density operator for the signal
state, of $0.901$. In fact, the departure from ideal purity is mainly due to
sidelobes (related to the sinc function) displaced from the central portion of
joint spectral intensity shown in Figs.~\ref{fig:estsime}(c) and (d). The
purity can be increased by filtering out these sidelobes, which in general
contain a small fraction of the total flux. Specifically, for the example
shown in Fig.~\ref{fig:estsime}, two separate narrowband rectangular-profile
spectral filters for the signal and idler modes with equal frequency bandwidth
of $9.35 \times10^{11} \mbox{rad }s^{-1}$ (corresponding to wavelength widths
$\Delta\lambda_{s} \approx0.1$nm and $\Delta\lambda_{i} \approx1.2$nm)
increases the purity to $0.981$, while reducing the flux by $\lesssim5\%$.

As pointed out already, one of the advantages of generating factorable,
symmetric photon pairs through SFWM rather than PDC in a second-order
nonlinear crystal, is that it presents greater (but still not arbitrary)
choice of the pump, signal and idler frequencies. Thus, for the example shown
above, we have chosen the pump so that it is compatible with, say, a
Ti:sapphire laser. One of the generated photons is centered at $1551$ nm,
which corresponds to the standard telecommunications band and could be
detected with an InGaAs-based avalanche photo-detector. The conjugate
generated photon is centered at $464$ nm, which can be detected with a
silicon-based avalanche photodiode (though with non-optimal efficiency, since
this wavelength is relatively far from the sensitivity peak of such
detectors). A natural application for such a source would be the heralding of
pure, single photons for transmission over standard optical telecommunication fibers.

\section{Co-polarized fields and non-degenerate pumps: symmetric factorable
states}

The use of non-degenerate pumps represents an additional degree of freedom,
not available for PDC in second-order crystals, which can be exploited for
state engineering purposes. The two pumps may differ both in terms of their
central frequencies ($\omega_{1}^{0}$ and $\omega_{2}^{0}$) and their
bandwidths ($\sigma_{1}$ and $\sigma_{2}$); in what follows we exploit both of
these aspects of non-degeneracy. The orientation of the phase-matching
function (see Eqs. (\ref{eq: ldelk}) and (\ref{PMF1})) in \{$\omega_{s}%
,\omega_{i}$\} space is determined by the angle $\theta_{si}%
=-\mbox{arctan}\left(  T_{s}/T_{i}\right)  $. Note that the orientation of the
phase-matching function depends on $\sigma_{1}$ and $\sigma_{2}$ (through
$T_{s}$ and $T_{i}$), while the shape of its profile and its width also depend
on $\sigma_{1}$ and $\sigma_{2}$ (through parameter $B$). A considerable
simplification results by imposing the condition $\sigma_{1} \ll\sigma_{2}$.
In this case, $T_{\mu}$ reduces to the corresponding degenerate-pump values
$\tau_{\mu}$ (with $\mu=s,i$). Furthermore, parameter $B$ reduces to
$1/(\sigma_{1} \tau_{p})$, so that the phase-matching function width exhibits
no dependence on the broader pump bandwidth ($\sigma_{2}$), while the
effective pump envelope function depends only on $\sigma_{2}$. This
de-coupling of $\sigma_{1}$ and $\sigma_{2}$ translates into a more
straightforward exploration of possible fiber geometries for the generation of
states with specific spectral entanglement properties.

\begin{figure}[t]
\begin{center}
\centering\includegraphics[width=11.5cm]{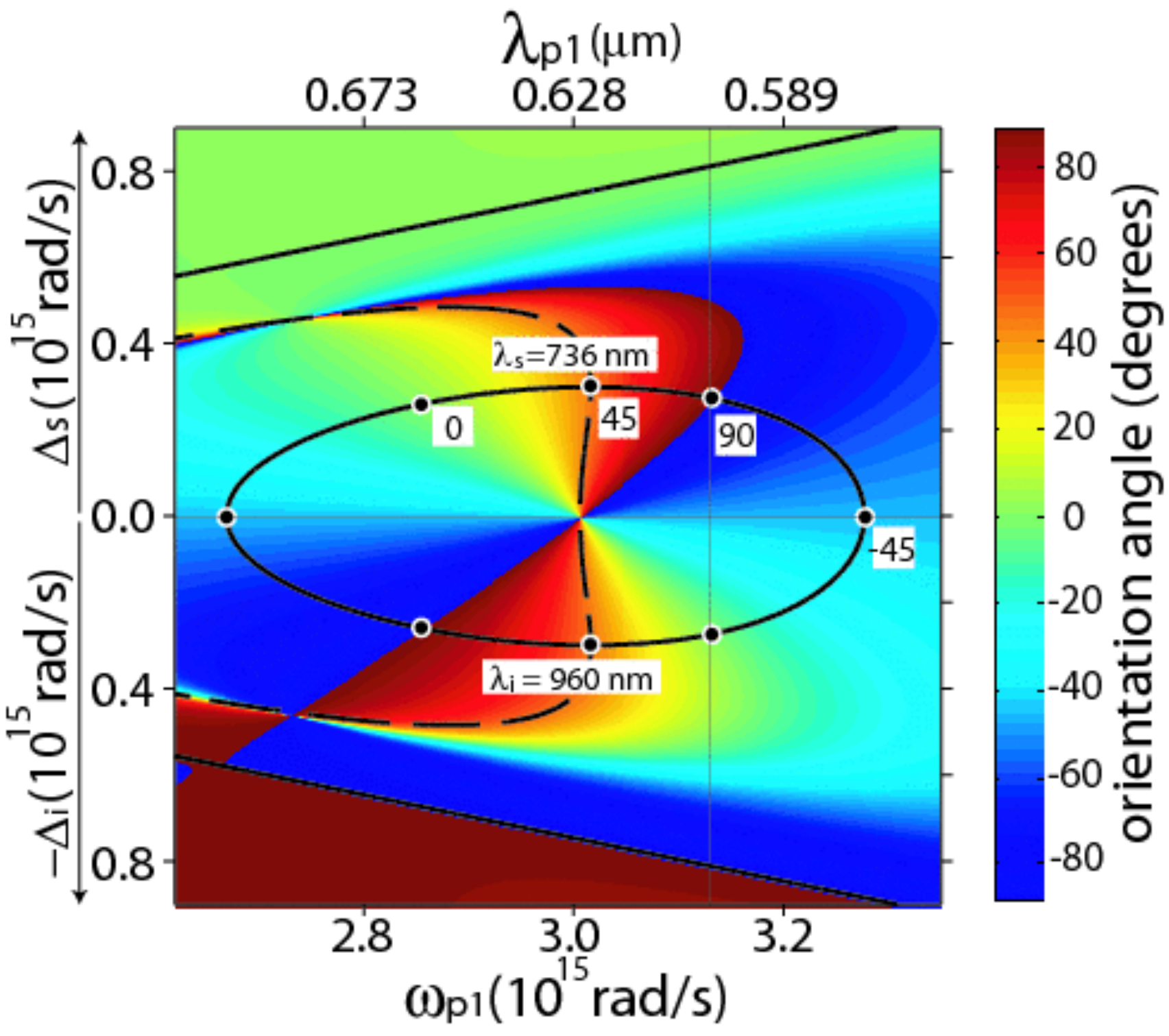}
\end{center}
\caption{Black, solid curve: phase-matching ($\Delta k=0$) contour for SFWM in
the non-degenerate pump regime. Colored background: phase-matching orientation
angle. Black, dotted line: frequencies that satisfy the group-velocity
matching condition (see Eq. (\ref{Eq:GVMcond})). Along the phase-matching
contour we have indicated particular angles of orientation of interest. }%
\label{cont_NDP}%
\end{figure}

A phase-matching diagram similar to that presented for degenerate pumps (see
Fig. \ref{fig: contordk}), may be prepared by maintaining one of the two pump
frequencies fixed (in this case pump $2$), where it is convenient to express
the generated frequencies as detunings from the \textit{mean} pump frequency
i.e. $\Delta_{s,i}=\omega_{s,i}-(\omega_{1}^{0}+\omega_{2}^{0})/2$; note that
energy conservation implies that $\Delta_{s}=-\Delta_{i}$. Figure
\ref{cont_NDP} illustrates the phase-matching properties for a specific
non-degenerate geometry, with $r=0.601\mu$m, $f=0.522$ and $\omega_{2}%
^{0}=1.508 \times10^{15}$rad/sec (that is, $\lambda_{2}^{0}=2 \pi c/\omega
_{2}^{0}=1250$ nm). The black solid curve represents the resulting
phase-matching contour. The two straight lines represent trivial
phase-matching branches (in which the created photons are degenerate with the
two pumps), while the loop represents the non-trivial phase-matching branch.
Depending on the fiber geometry, the non-trivial branch may become large
enough to overlap the trivial branches, or may shrink down to a single point.
For the former case, if self-phase modulation becomes appreciable, this
contour splits into three distinct loops (rather than two as for degenerate
pumps). The colored background represents the orientation angle $\theta_{si}$
ranging from $-90^{\circ}$ in blue to $90^{\circ}$ in red.


A factorable, symmetric two-photon state is produced if i) $\theta
_{si}=45^{\circ}$ which implies $T_{s}=-T_{i}$ and ii) the spectral widths of
the phase-matching function $\phi(\omega_{s},\omega_{i})$ and the pump
envelope function $\alpha(\omega_{s},\omega_{i})$ are equal to each other.
From Eqs. (\ref{eq: taus}) it is straightforward to show that the first
condition is satisfied if the four fields satisfy%

\begin{equation}
2k_{2}^{(1)}-k_{s}^{(1)}-k_{i}^{(1)}=-2(k_{1}^{(1)}-k_{2}^{(1)}) \sigma
_{1}^{2}/(\sigma_{1}^{2}+\sigma_{2}^{2}). \label{Eq:GVMcond}%
\end{equation}

This represents a generalization of the equivalent group-velocity matching
condition for degenerate pumps (see Eq.~\ref{eq: CoDP_i}, and the text which
follows). If $\sigma_{1}\ll\sigma_{2}$, the right-hand side of Eq.
(\ref{Eq:GVMcond}) vanishes, leading to a condition that is identical in form
to that obtained for degenerate pumps (where the broader pump now plays the
role of the degenerate pump). In Fig. \ref{cont_NDP}, pairs $\{\Delta
_{s},\omega_{p1}\}$ that satisfy this generalized group-velocity matching
condition are represented by a dashed line. Note that the position and shape
of this contour in general depend on $\sigma_{1}$ and $\sigma_{2}$; however,
in the case $\sigma_{1}\ll\sigma_{2}$, the contour becomes decoupled from
these bandwidths. Intersection points between the two contours determine the
center frequency for pump field $1$ that satisfies the requisite
group-velocity matching, in addition to phase-matching. The values of $r$ and
$f$ are selected so that $\theta_{si}=45^{\circ}$ is satisfied \textit{and} so
that the broadband pump frequency $\omega_{1}^{0}$ takes a certain desired
value (in this case, $\lambda_{1}^{0}=2 \pi c/\omega_{1}^{0}=0.625\mu$m). This
leads to the generated signal and idler wavelengths $736$nm and $960$ nm. The
fiber length, along with the two bandwidths (subject to $\sigma_{1}\ll
\sigma_{2}$) are selected so as to match the widths of the pump envelope and
phasematching functions. For the specific geometry shown, these values are
$L=25$cm, while $\sigma_{1}$ and $\sigma_{2}$ are given in terms of the
corresponding FWHM bandwidths: $\Delta\lambda_{1}=1.51$ nm and $\Delta
\lambda_{2}=0.12$ nm.

\begin{figure}[t]
\begin{center}
\centering\includegraphics[width=11.5cm]{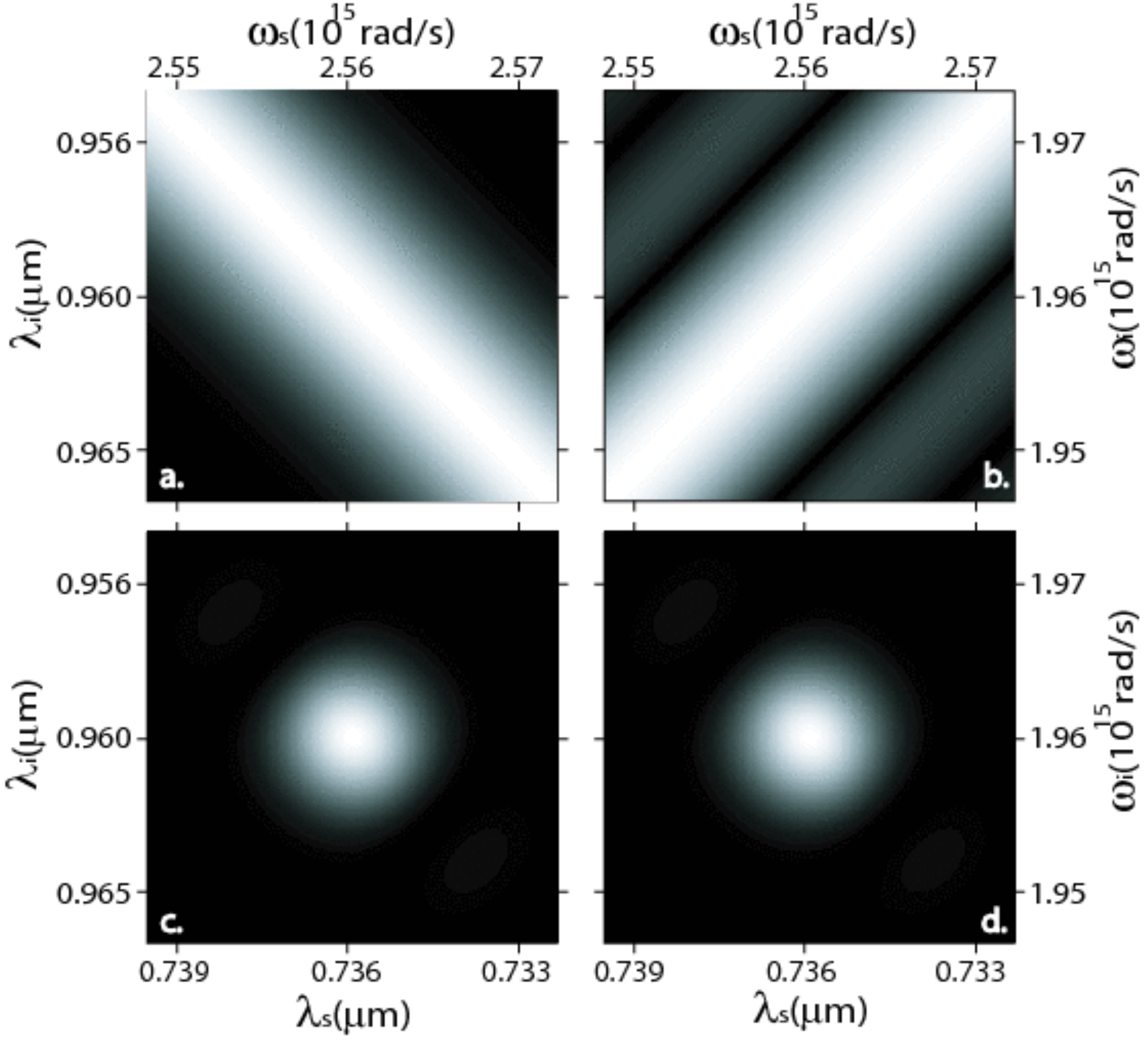}
\end{center}
\caption{Joint spectral intensity (JSI) $|F(\omega_{s},\omega_{i})|^{2}$
obtained for the fiber geometry assumed for Fig. \ref{cont_NDP}. (a) Pump
envelope function $\alpha(\omega_{s},\omega_{i})$. (b) Phase-matching function
$\phi(\omega_{s},\omega_{i})$. (c) Analytic JSI. (d) JSI obtained by numerical
integration of Eq. (\ref{eq: JSA}). }%
\label{NDPsim}%
\end{figure}

An important motivation for exploiting pump non-degeneracy is that it permits
SFWM geometries with pumps which are sufficiently distinct from the signal and
idler photons, and yet where the latter are in relative proximity. This is
important on the one hand for the suppression of contamination from
spontaneous Raman scattering and on the other hand to ensure that both
generated photons can be detected with available efficient single-photon
detectors. The pump wavelengths were selected so that they can be obtained
from a modelocked Cr:Forsterite laser and its second harmonic, while the
generated light is within the detection bandwidth of silicon avalanche
photo-diodes. Figure \ref{NDPsim}(a) shows the pump envelope function
$\alpha(\omega_{s},\omega_{i})$. Figure \ref{NDPsim}(b) shows the
phase-matching function $\phi\left(  \omega_{s},\omega_{i}\right)  $ (see
Eq.(\ref{PMF1})) for which $B=1.73$. Figure \ref{NDPsim}(c) shows the
resulting joint spectral intensity which exhibits a factorable character,
obtained using the linear-dispersion approximation (see Eq. (\ref{eq: ldelk}%
)). For comparison, Fig. \ref{NDPsim}(d) shows the joint spectral intensity
obtained by numerical integration of Eq.(\ref{eq: JSA}), exhibiting excellent
agreement with Fig. \ref{NDPsim}(c). This source leads to a
numerically-obtained state purity of $0.89$. As in the case of a factorable
symmetric state obtained with degenerate pumps (see Fig.~\ref{fig:estsime}),
the purity can be increased by filtering out the sidelobes in the joint
spectral intensity at small cost in terms of collected flux. Two separate
narrowband rectangular-profile spectral filters for the signal and idler modes
with equal frequency bandwidth of $12.90 \times10^{12} \mbox{rad }s^{-1}$
(corresponding to wavelength widths $\Delta\lambda_{s} \approx3.7$nm and
$\Delta\lambda_{i} \approx6.3$nm) increases the purity to $0.98$, while
reducing the flux by $\lesssim5.9\%$.


\section{Cross-polarized fields and degenerate pumps: asymmetric factorable
states}

Birefringence is an additional degree of freedom that we can use to satisfy
the conditions necessary for generating factorable two-photon states. Typical
hexagonal pattern PCF can exhibit significant birefringence due to slight
asymmetries in the hole pattern or internal stresses in the glass.
Birefringence can also be purposefully added to the structure through defects
such as two larger holes or elliptical cores. In PCFs, the birefringence
typically ranges from $\Delta n\approx$ $1\times10^{-5}$, for normal PCF, to
$\Delta n\approx$ $7\times10^{-3}$ \cite{Ortigosa-Blanch2004} for polarization
maintaining PCFs.

Conservation of angular momentum restricts the types of FWM processes that can
occur in a fiber, with the result that the polarizations of the four fields
must appear in pairs in the process \cite{agrawal2007, berlhoer1970}. It
follows that there are three distinct pair generation processes that can occur
in fibers: $xx\rightarrow xx$, $xy\rightarrow xy$, $xx\rightarrow yy$, where
$x$ and $y$ can be any two linear orthogonal polarizations. Photon pair
generation with frequency separated pumps in the $xy\rightarrow xy$ process
was analyzed in 2004 \cite{McKinstrie04} and demonstrated \cite{fan2005} in
2005.
In contrast, we limit our discussion to a single pump pulse polarized along
one fiber axis (e.g. $xx\rightarrow yy$, $2\omega_{p}\rightarrow\omega
_{i}+\omega_{s}$). The first realization of birefringent FWM was a
demonstration of this process in regular fibers \cite{stolen1981}. It was
demonstrated in PCFs in 2006 \cite{Kruhlak2006}.

Single-photon detectors operating in the visible region of the
spectrum are generally less costly and have higher quantum
efficiencies and lower noise than single-photon detectors
operating in the near-infrared region. Therefore for some
applications it is desirable for both photons from a SFWM pair
to be produced in the visible region. This limits the
usefulness for these applications of PCFs which have
outer-branch solutions far separated from the pump frequency,
as shown in Fig. \ref{fig: contordk}(a), as one of the photons
is generated in the infrared. On the other hand, the photons
should be at least 50 THz from the pump to reduce background
from spontaneous Raman scattering. This limits the usefulness
of the inner branch solutions, for which the generated photons
are typically less then 5 THZ from the pump
\cite{fiorentino02}. The benefit of birefringence is that it
creates new factorable solutions inside the visible region yet
still separated from the pump by more than the Raman bandwidth.
Unlike the nondegenerate case, it accomplishes this without the
experimental complexity of two pumps. This can
be seen by considering the relevant phase-matching equation,%

\begin{align}
\Delta k\left(  \omega_{p},\Delta_{s}\right)   &  =2k_{x}\left(  \omega
_{p}\right)  -k_{y}\left(  \omega_{p}+\Delta_{s}\right)  -k_{y}\left(
\omega_{p}-\Delta_{s}\right)  -\frac{2}{3}\gamma P\label{biPM}\\
&  \approx2k_{x}\left(  \omega_{p}\right)  -k_{x}\left(  \omega_{p}+\Delta
_{s}\right)  -k_{x}\left(  \omega_{p}-\Delta_{s}\right)  +2\Delta
n\frac{\omega_{p}}{c}-\frac{2}{3}\gamma P, \label{approxbiPM}%
\end{align}

\noindent where the power-induced term is 1/3 of that in the co-polarized case
\cite{agrawal2007}. Some insight can be gained by separating the birefringent
contribution to $\Delta k$ from the individual wavevectors as in Eq.
(\ref{approxbiPM}). It was experimentally observed in \cite{Kruhlak2006} that
$\Delta n$ is approximately independent of frequency and we note that
$\omega_{p}$ is relatively constant in comparison to $k_{x}\left(  \omega
_{p}\right)  $. Consequently, the birefringent contribution to the wavevector
mismatch functions much like the power phase-modulation term in Eq.
(\ref{eq: delk}), introducing in the $\Delta k=0$ curve a splitting of the
$\Delta_{s}=0$ solution (creating the inner branch in Fig. \ref{fig: contordk}%
). Unlike the power term, the birefringence can create a large splitting and
can be either positive or negative.


\begin{figure}[t]
\begin{center}
\centering\includegraphics[width=12cm]{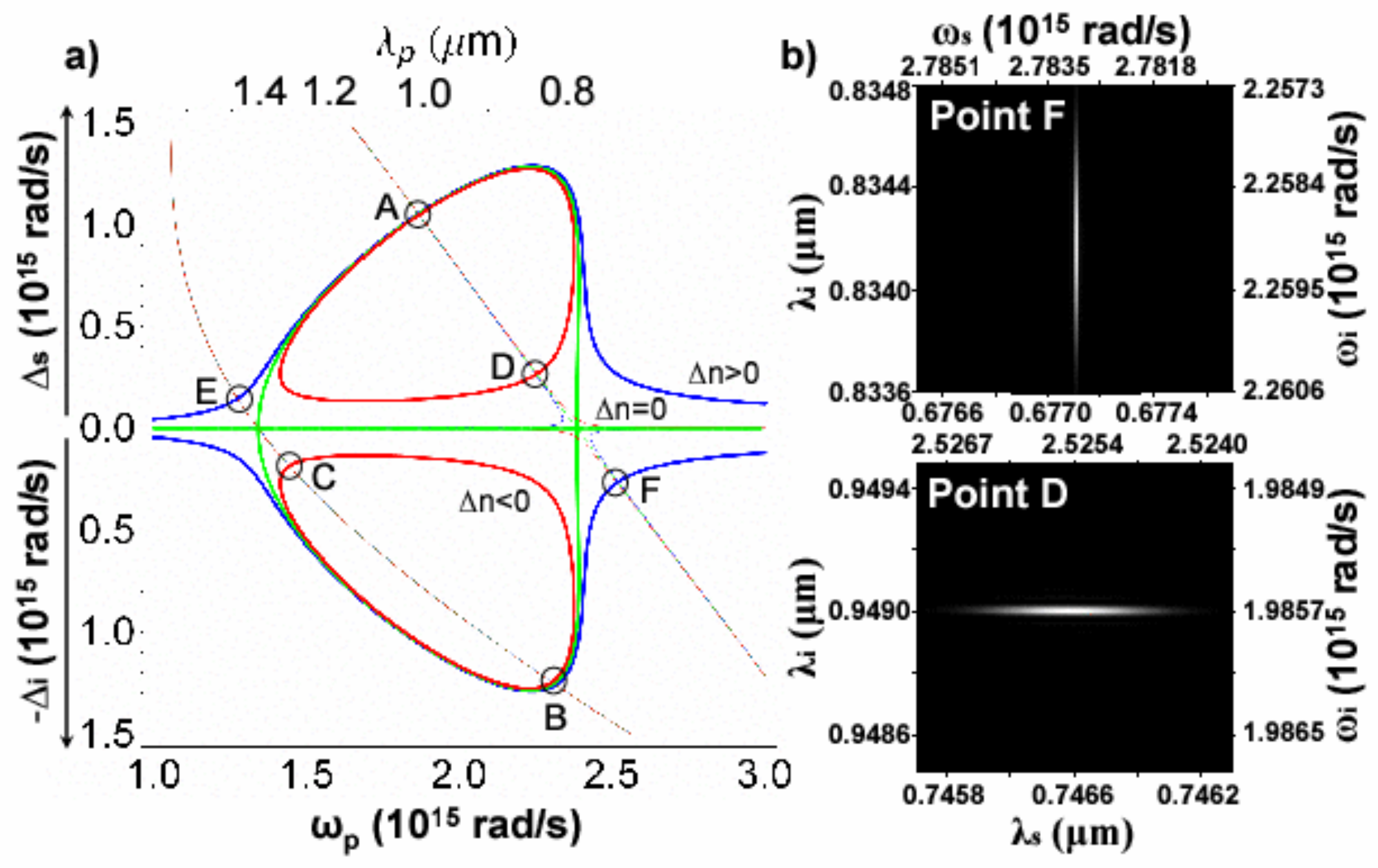}
\end{center}
\par
\caption{a) Cross-polarization phase-matching $\Delta k=0$ (thick) and
group-velocity matching $T_{s,i}=0$ (thin) curves ( $f=0.43$, $d=1.75\mu m, $
ZDWs $=790$ nm, $1404$ nm -- similar to NL-1.7-790 from Crystal Fiber). With
power $P=0$, three pairs of curves are plotted with $\Delta d=-0.001,0,0.001$
$\mu m,$ respectively resulting in a birefringence of $\Delta n=-3\times
10^{-5},0,3\times10^{-5}$ (red, green, blue). Points A through F bound regions
in which factorability is possible. b) Full dispersion numerical JSIs for the
asymmetric states corresponding to points D ($\theta_{si}=0^{\circ}%
,\lambda_{s}$ $=746.4$ nm and $\lambda_{i}$ $=949.0$ nm, $\Delta\lambda_{p}=$
$0.30$ nm, and $L=30$ m) and F ($\theta_{si}=90^{\circ}$, $\lambda_{s}$
$=677.1$ nm, $\lambda_{i}$ $=$ $834.2$ nm, $\Delta\lambda_{p}=$ $0.30$ nm, and
$L=30$ m).}%
\label{birefring}%
\end{figure}


We begin by extending the step index model reviewed in Section 3 to
birefringent fiber. In this extension, each transverse axis of the fiber is
modeled as a separate fiber. The axes have the same air-filling fraction
parameter $f$ but different core diameters $d$ and $d+\Delta d,$ simulating
the birefringence \cite{Wong2005, Kruhlak2006}. We use this model for all the
figures in this section. We also neglect the power term in Eq. (\ref{biPM})
since it is typically much smaller than the birefringent contribution.
Although this model is accurate only for small birefringence where the fiber
asymmetry is small, it allows us to predict universal properties of the SFWM
across a broad range of configurations.

In Fig. \ref{birefring}, the phase-matching ($\Delta k\left(  \omega
_{p},\Delta_{s,i}\right)  =0$) and group-velocity matching ($T_{s,i}=0$)
curves are plotted for a PCF fiber with ZDWs at 790 nm and 1404 nm, where
$d=1.75 \mu$m and $f=0.43$. Three pairs of curves are plotted for $\Delta
d=-0.001,0,0.001$ $\mu$\textrm{m}, resulting in a birefringence of $\Delta
n=-3\times10^{-5},0,3\times10^{-5}$ (where $\Delta n=n_{y}-n_{x}$). The
corresponding shift in the ZDW is -0.07, 0, 0.07 nm and $k^{(1)}_{x} -
k^{(1)}_{y} = -0.5, 0, 0.5$ ps/km, respectively. The segments of the $\Delta
k=0$ curve bounded by intersections with the $T_{s,i}=0$ curve are regions
where factorability is possible. For both positive and negative $\Delta n$ the
outer branch intersections remain approximately the same as in the copolarized
degenerate case. In addition, negative $\Delta n$ creates two intersections
(Points C and D) in between the two ZDWs. In between these points
factorability is possible. Likewise, positive $\Delta n$ also creates two
intersections (Points E and F). However, now it is outside the region bounded
by the intersections that factorability is possible. The splitting of the
phase-matching ($\Delta k = 0$) solution from the $\Delta_{s,i}=0$ line in the
$\Delta n<0$ case is similar to that from the power-induced phase modulation
term $2\gamma P$ in the phase matching equation Eq. (\ref{eq: delk}), but is
an order of magnitude larger. Unlike power-induced inner branch FWM,
birefringence allows for phase-matching beyond the Raman peak and is
relatively insensitive to pump laser power fluctuations. In contrast, the
$\Delta n>0$ case is qualitatively different from either of these since
$2\gamma P$ is always positive in silica. Moreover, it allows for factorable
state generation over an unprecedented pump wavelength range, essentially
anywhere outside the ZDWs.


\begin{figure}[t]
\begin{center}
\includegraphics[width=12cm]{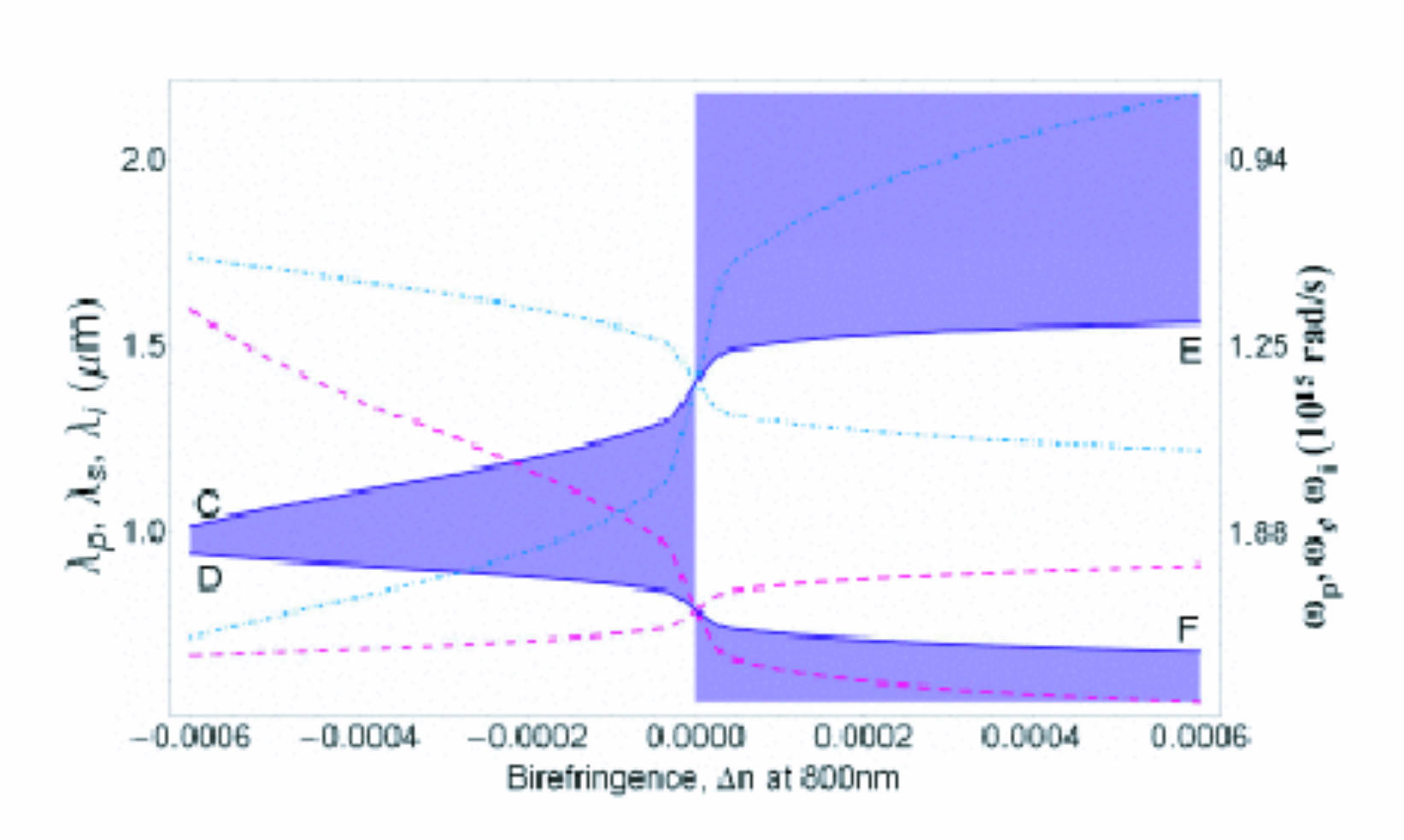}
\end{center}
\par
\caption{Points C through F from Fig. \ref{birefring} are plotted as a
function of the birefringence $\Delta n.$ Thus, the solid dark blue curves
give the pump wavelength $\lambda_{p}$ of asymmetric factorable solutions. The
corresponding idler and signal wavelengths $\lambda_{i}$ and $\lambda_{s}$ are
given by the magenta dashed curves (for points D and F) and light-blue
dash-dotted curves (for points C and E). The shaded regions indicate the range
in which factorable states can be generated provide the pump bandwidth is
matched to the fiber length. The fiber parameters are the same as in Fig.
\ref{birefring}.}%
\label{gvmsol}%
\end{figure}


Figure \ref{birefring}(b) shows the joint spectral intensity in two
cases that lead to high degrees of factorability. Factorability
corresponds in both cases to the absence of correlation between
signal and idler frequencies, as can be seen in the shape of the
JSI. In either case, if the idler photon is detected (without
spectral filtering), then a nearly pure-state signal photon is
heralded. Numerically calculated heralded photon purity for the
cases at point F and point D are 0.988 and 0.993, respectively.

In Fig. \ref{gvmsol}, the intersection of the $\Delta k=0$ and $T_{s,i}=0$
curves is plotted as a function of the $\Delta n$ at 800 nm. Thus for each
$\Delta n$ the figure indicates the pump wavelength $\lambda_{p}$ that creates
an asymmetric factorable state ($\theta_{si}=0^{\circ}$ or $\theta
_{si}=90^{\circ}$), as well as the center wavelengths $\lambda_{s}$ and
$\lambda_{i}$ of the generated photons. At these points, the pump bandwidth
must satisfy condition $T_{s,i}>>\sigma^{-1}$ but is otherwise arbitrary. The
shaded regions indicate the $\lambda_{p}$ range in which factorability is
possible (including the symmetric state) if the pump bandwidth is set
according to Eq. (\ref{eq: CoDP_ii}).

It has been pointed out in the FWM literature that birefringent FWM has the
added benefit of reducing Raman contamination in the signal and idler
\cite{murdoch1995}. Recently this has also been forwarded \ as a strategy to
reduce background for pair-generation \cite{Lin2006}. Raman gain $g_{Raman}$
is reduced by as much as an order of magnitude in the axis orthogonal to the
strong pump pulse \cite{stolen2003}. However, the relevant $\chi^{(3)}$ for
birefringent pair production is also reduced by a factor of three compared to
co-polarized FWM \cite{agrawal2007}. Tripling the power $P$ to compensate,
Ref. \cite{Lin2007} studied the pair-correlation produced by the fiber,
$\rho_{c}(0)=\frac{\left\langle I_{i}I_{s}\right\rangle }{\left\langle
I_{i}\right\rangle \left\langle I_{i}\right\rangle }-1,$ where $I$ is the
intensity of signal or idler mode. They found that at the Raman peak,
$\rho_{c}(0)$ was 7 for co-polarized SFWM as compared to 60 for
cross-polarized SFWM, indicating that the cross-polarized case can be a higher
quality photon-pair source.

\section{Co-polarized fields: ultra-broadband two-photon states}

The flexibility afforded by PCFs in engineering the spectral
correlations of photon pairs permits the generation of a wide
class of states. For instance, this may be illustrated by
considering the generation of quantum states in which the
photon pair component has an extraordinarily high degree of
frequency anti-correlation\cite{footnote}. The photons
individually have an ultrabroad bandwidth, yet are very tightly
anticorrelated in frequency, and highly correlated in time (to
the level of a single optical cycle). We have identified
conditions for which ultra-broadband two-photon states can be
generated in the degenerate pump regime by SFWM. Such
ultrabroad states have recently been generated by PDC
\cite{odonnell07}. As pointed out above, for fiber geometries
where the phase-matching diagram occurs in the form of loops
(which is true for most cases of interest), it is possible to
obtain an arbitrary phase-matching orientation on
$\{\omega_{s},\omega_{i}\}$ space. By imposing constraints on
higher dispersive orders, it becomes possible to generate
states with yet different character. In particular, in analogy
with the generation of ultra-broadband two-photon states via
PDC~\cite{odonnell07,zhang07}, it can be shown that in the case
of SFWM, the bandwidth of the generated light can be made
particularly large if the fourth-order $k^{(4)}$ and second
order $k^{(2)}$ coefficients vanish simultaneously. In this
case, a degenerate pump at the zero-group-velocity-dispersion
frequency can lead to a state with a remarkably broad bandwidth
(hundreds of nanometers), even for a nearly monochromatic pump.
For example, for the PCF described previously with
$r=0.616\mu$m and $f=0.6$, whose phase-matching curve is shown
in Fig. \ref{fig: contordk}, pumping at a frequency around
$2.82 \times10^{15}$ rad/sec (wavelength $\lambda= 668$ nm),
where the curve is nearly vertical, would create photons with a
bandwidth of around $2 \times10^{15}$ rad/sec ($\Delta\lambda
\approx542$ nm) centered at the pump frequency, depending on
the bandwidth of the pump. Previously, light produced by
co-polarized pumps was shown to have a bandwidth of 50 nm for
CW pumps \cite{jiang07} and 100 nm for pulsed pumps
\cite{chavez07}. For cross-polarized CW pumps broadband light
of 34 nm was produced \cite{radic03}.


\section{Conclusions}

We have studied theoretically the spontaneous four wave mixing
process in single mode photonic crystal fiber. We have shown
that it is possible to design a source so as to yield photon
pairs with a broad class of spectrally engineered properties,
including factorable and ultrabroadband states. These results
are achieved by adjusting the source configuration, including
the dispersion (determined by the core radius and air filling
fraction) and birefringence of the fiber and the pump frequency
or frequencies, bandwidth(s) and polarization(s). The
flexibility of the two-photon state engineering arises from the
interplay of this large number of control parameters.
Factorable signal-idler pair generation allows the production
of heralded, high purity single-photon wavepackets, with
minimal or no need for spectral filtering.

Factorable state generation is possible if certain group-velocity matching
conditions are satisfied, in addition to standard phase-matching. Most fibers
(all those for which the phase-matching curve occurs in loops) can achieve all
appropriate group-velocity matching conditions by suitable choice of the pump
wavelength. This allows one to realize all possible orientations of the joint
spectral amplitude of the generated photon pairs. The size of the
phase-matching loop is largely determined by the dispersion of the fiber,
hence the utility of PCFs. Therefore we find that for specific choices of pump
frequency most fiber designs can yield factorable states. This is in contrast
with parametric downconversion in second-order nonlinear crystals where such
conditions are met at best at isolated frequencies for certain materials, or
using more complicated microstructures \cite{raymer05,uren06}.

An important design consideration is the possible contamination of the
frequency sideband modes by photons from spontaneous Raman scattering. This
problem can be circumvented in part by cooling the fiber to cryogenic
temperatures, thereby suppressing the Raman scattering \cite{li04,takesue05}.
We have shown how to circumvent the problem by exploiting a phase-matching
configuration where the signal and the idler frequencies are widely separated
from the pump frequency (see Fig. \ref{fig: contordk}). If desired, the phase
matching in non-birefringent fiber can be engineered to create photon pairs
having one photon in the infrared and one photon in the visible. Alternative
approaches also discussed in this paper are cross-polarized FWM in
birefringent fiber, where the pump and generated photons have orthogonal
polarizations, and the use of non-degenerate pumps in a co-polarized regime.
We have shown that these approaches lead to sources capable of generating
factorable photon pairs where the signal and idler photons are both in the
visible, and at the same time are both sufficiently separated from the pump
frequency to avoid Raman contamination. We expect these results to be useful
in the design of sources for practical implementations of quantum information
processing technologies.

\section*{Acknowledgments}

KGP, and RRR acknowledge CONACYT-M\'exico for partial funding through project
no. 46492, and ABU acknowledges support from CONACYT-Mexico through grant
46370. IAW, JL and OC were supported by the EPSRC (UK) through the QIP IRC
(GR/S82716/01) and through project EP/C013840/1 and by the European Commission
under the Integrated Project Qubit Applications (QAP) funded by the IST
directorate as Contract Number 015848. HJM and MGR were supported by the
National Science Foundation grants numbers ECS-0621723 and PHY-0554842.

\begin{thebibliography}{99}
\bibitem {footnote}Such a state is typically referred to as highly entangled,
but one should keep in mind that the large vacuum component of the state
renders this ``entanglement'' useful only in a post-selection experiment.
\end{thebibliography}
\end{document}